\documentclass[journal=jpclcd,manuscript=letter]{achemso}%
\usepackage[version=3]{mhchem} 
\usepackage[T1]{fontenc}       
\usepackage{caption}
\usepackage{booktabs}
\usepackage[utf8]{inputenc}

\author{Antoine Comby}
\affiliation[Bordeaux]
{Universit\'e de Bordeaux-CNRS-CEA, CELIA, UMR5107, F33405 Talence, France}
\author{Samuel Beaulieu}
\affiliation[Bordeaux]
{Universit\'e de Bordeaux-CNRS-CEA, CELIA, UMR5107, F33405 Talence, France}
\alsoaffiliation[Canada]{Institut National de la Recherche Scientifique, Centre \'EMT, Varennes, Quebec, Canada}
\author{Martial Boggio-Pasqua}
\affiliation[Toulouse]{Universit\'e de Toulouse, CNRS,LCPQ-IRSAMC, UMR 5626, 118 route de Narbonne, 31062 Toulouse, France.}
\author{Dominique Descamps}
\affiliation[Bordeaux]
{Universit\'e de Bordeaux-CNRS-CEA, CELIA, UMR5107, F33405 Talence, France}
\author{Francois L\'egar\'e}
\affiliation[Canada]{Institut National de la Recherche Scientifique, Centre \'EMT, Varennes, Quebec, Canada}
\author{Laurent Nahon}
\affiliation[SOLEIL]{Synchrotron SOLEIL, L'Orme des Merisiers, Saint Aubin BP 48, 91192, Gif sur Yvette Cedex, France}
\author{St\'ephane Petit}
\affiliation[Bordeaux]
{Universit\'e de Bordeaux-CNRS-CEA, CELIA, UMR5107, F33405 Talence, France}
\author{Bernard Pons}
\affiliation[Bordeaux]
{Universit\'e de Bordeaux-CNRS-CEA, CELIA, UMR5107, F33405 Talence, France}
\author{Baptiste Fabre}
\affiliation[Bordeaux]
{Universit\'e de Bordeaux-CNRS-CEA, CELIA, UMR5107, F33405 Talence, France}
\author{Yann Mairesse}
\affiliation[Bordeaux]
{Universit\'e de Bordeaux-CNRS-CEA, CELIA, UMR5107, F33405 Talence, France}
\author{Val\'erie Blanchet}
\affiliation[Bordeaux]
{Universit\'e de Bordeaux-CNRS-CEA, CELIA, UMR5107, F33405 Talence, France}
\email{valerie.blanchet@celia.u-bordeaux.fr}

\title[An \textsf{achemso} demo]
  {Relaxation Dynamics in Photoexcited Chiral Molecules Studied by Time-Resolved Photoelectron Circular Dichroism: Towards Chiral Femtochemistry}

\abbreviations{Femtochemistry, Chirality, Vibronic relaxation, Rydberg states, Fenchone}

\begin{document}

\begin{abstract}
  Unravelling the main initial dynamics responsible for chiral recognition is a key step in the understanding of many biological processes. However this challenging task requires a sensitive enantiospecific probe to investigate molecular dynamics on their natural femtosecond timescale. Here we show that, in the gas phase, the ultrafast relaxation dynamics of photoexcited chiral molecules can be tracked by recording Time-Resolved PhotoElectron Circular Dichroism (TR-PECD) resulting from the photoionisation by a circularly polarized probe pulse. A large forward/backward asymmetry along the probe propagation axis is observed in the photoelectron angular distribution. Its evolution with pump-probe delay reveals ultrafast dynamics that are inaccessible in the angle-integrated photoelectron spectrum nor via the usual electron emission anisotropy parameter ($\beta$). PECD, which originates from the electron scattering in the chiral molecular potential, appears as a new sensitive observable for ultrafast molecular dynamics in chiral systems.
\end{abstract}


Chiral species exhibit enantio-specific properties when they are embedded into a chiral medium. For instance, chiral molecules interact with each other to form chiral complexes via the so-called chiral recognition process, this `molecular handshake' being at the basis of stereochemistry.\cite{zehnacker_chirality_2008} Enantio-specific properties also arise from the interaction of chiral molecules with `chiral light', such as Circularly Polarized Light (CPL), leading to various types of Circular Dichroism (CD).\cite{berova_circular_2000} As calculated\cite{ritchie_theoretical_1975,powis_photoelectron_2000} and experimentally-demonstrated,\cite{bowering_asymmetry_2001} CPL-induced photoionisation of randomly-oriented pure enantiomers gives rise to a forward/backward asymmetry of photoelectron emission along the photon propagation axis \textbf{\textit{k}}, referred to as PhotoElectron Circular Dichroism (PECD). Based upon a differential measurement, this chiroptical effect can be fully described in the pure electric dipole approximation, leading to very strong asymmetries. The outgoing electron scatters in the mean chiral molecular potential involving all the other electrons – mostly the valence ones, which are the main actors of chemical reactivity.  This electron scattering occurs in a space whose orientation is referenced by the circularly polarized ionising field. Despite the random-orientation of the molecules, the resulting photoelectron angular distribution shows an intense forward/backward asymmetry. The magnitude of this asymmetry, determined by the interferences between the different partial waves describing the outgoing electron wavepacket, can reach the order of several percent. The sign of this asymmetry reverses with the handedness of the enantiomer ($e$) or the helicity of the ionising light ($p$). PECD is an orbital-specific effect that also depends on the final state; i.e. the photoelectron continuum.  It is very sensitive to permanent – isomers, conformers, clustering – and dynamical – vibrational excitation – molecular structures.\cite{nahon_valence_2015} As such, PECD is a useful observable to investigate dynamics in chiral molecules. So far, experiments, mostly synchrotron radiation based, have focused on the photoionisation of molecules from their ground electronic state (S$_0$) in its equilibrium geometry. Recently, a laser-based one-color Resonant Enhanced Multiphoton ionisation (REMPI)-PECD scheme involving several intermediate states has been demonstrated, however without any time resolution.\cite{lux_circular_2012,lehmann_imaging_2013,Beaulieu_2016}  The possibility of observing PECD from excited states, for instance loosely bound electronic states, remains an open question. 
The present paper focuses on the investigation of PECD from an electronic excited state to establish the sensitivity of PECD as a probe of the ultrafast relaxation taking place in this state. To record Time-Resolved PECD (TR-PECD), a linearly polarized femtosecond pump pulse photoexcites chiral molecules and a time-delayed circularly polarized femtosecond probe pulse photoionises them. Key features of this two-color two-photon experiment are schematised in Figure 1.

\begin{figure}
\includegraphics{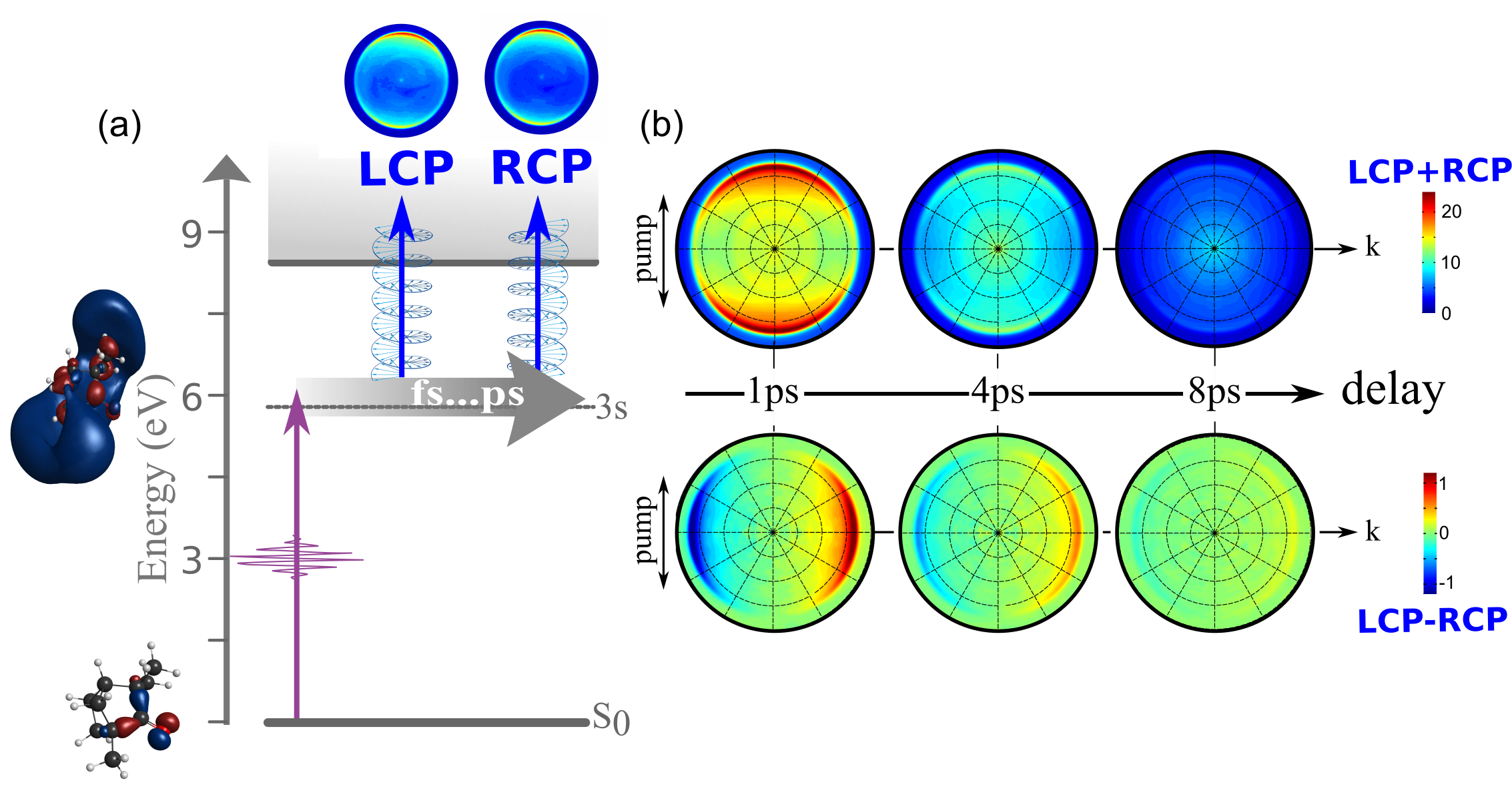}
\caption{(a) Schematic of the excitation scheme used to measure TR-PECD from the 3s Rydberg state of fenchone. The HOMO and the 3s orbitals are drawn to illustrate the difference in the initial wavefunction of the outgoing electron. (b) The photoelectron images recorded on (1R)-(-)-fenchone ($e=-1$) for each probe helicity ($LCP$-$\,p=1$ or $RCP$-$\,p=-1$) are subtracted one from the other $(LCP-RCP)$ to extract the unnormalised odd Legendre polynomial coefficients, while the sum of these images $(LCP+RCP)$ provides the photoelectron spectrum (PES). These $(LCP-RCP)$ and $(LCP+RCP)$ images, respectively called `difference'- and `sum'-images, are shown for three pump-probe delays. Note that the two photon beams are quasi co-propagating with polarization axis parallel to the detector (see SI).}
\end{figure} 

 The target molecule we investigate here is the fenchone (C$_{10}$H$_{16}$O), a bicyclic ketone whose rigid structure ensures the existence of a single conformer. Fenchone has been extensively studied by VUV-PECD with and without coincidence detection,\cite{nahon_determination_2016,ferre_table-top_2015} by multiphoton\cite{lux_photoelectron_2015}, above threshold or tunnel ionisation\cite{Beaulieu_2016} PECD as well as by absorption circular dichroism.\cite{pulm_theoretical_1997} However, none of these experiments was time-resolved, and all of them probed a molecule in its ground electronic state defined by its equilibrium geometry, with a possible contribution from the excited states in the REMPI-PECD experiments. The absorption spectrum of fenchone is characterized by a weak valence band around 4.6 eV, followed by strong Rydberg bands starting around 6 eV.\cite{pulm_theoretical_1997}  We chose to excite the molecules to their first Rydberg state (3s)  in order to drastically change the initial orbital of the outgoing electron compared to the ground state, as illustrated in Figure 1. The pump pulse spectrum is centred at 6.17 eV with a 30 meV full width half maximum bandwidth (201 nm, 80 fs duration if Fourier-transform-limited). The probe (403 nm, 35 meV with 70 fs duration) is chosen to photoionise the system $\sim$0.6 eV above the vertical ionisation threshold (IP$_v$=8.72 eV), in the energy range where the largest VUV PECD has been measured (15\%).\cite{nahon_determination_2016} Since the pump pulse is linearly polarized, the sensitivity to chirality only comes into play in the ionisation step, via the helicity of the probe pulse. \\
The photoelectron images are collected by a velocity map imaging detector for alternating probe helicities at each pump-probe delay, as described in the Experimental Methods section and the S.I. The momentum distribution of the photoelectrons (angular resolved photoemission spectrum $ARPES$) for a probe helicity $p=\pm 1$ (Left/Right) and one given enantiomer ($e$) is defined by a function $ARPES^{(p,e)}(\varphi,\theta,E_{kin})$ with $\theta$ the photoelectron ejection angle with respect to the probe propagation axis \textbf{\textit{k}}, $\varphi$ its azimuthal angle and $E_{kin}$ its kinetic energy. When a cylindrical symmetry in the emission angle $\varphi$ of the photoelectron is achieved along an axis parallel to the detector, the $ARPES^{(p,e)}(\theta,E_{kin})$ distribution is a simple sum of Legendre polynomials $[B^{(p,e)}_i(E_{kin}) P_i(\cos(\theta))]$ and can be retrieved from its VMI 2D-projection $PROJ^{(p,e)}(\theta,E_{kin})$ by using a p-Basex analysis for instance.\cite{garcia_two-dimensional_2004}\\
As the $P_{even}$ Legendre polynomials are symmetric along the optical axis \textbf{\textit{k}}, the $B_{even}$ coefficients encode the angular distribution of the standard photoelectron image: the PES is the angle-integrated photoelectron intensity $B_0(E_{kin})$ and the anisotropy parameter $b_2(E_{kin})=-\beta/2$ (with $\beta$, the usual anisotropy parameter retrieved for a linearly polarized photoionisation). These even parameters do not depend on the laser helicity ($p$) or enantiomer ($e$). The $P_{odd}$ Legendre polynomials are antisymmetric along \textbf{\textit{k}}, consequently the $B_{odd}^{(p,e)}(E_{kin})$ coefficients are non-zero only for chiral molecules photoionised by elliptically polarized light ($p\neq 0$) and encode the PECD. In a 2-photon ionisation, the sum  of Legendre polynomials is expected to go up to $P_4 (\cos(\theta))$, with a drastically reduced $B_4$  in a (1+1') resonant ionisation. \\
In the present experiment, the linear pump polarization as well as the circular probe polarization are defined by quantification axes perpendicular to each other and both parallel to the detector axis. The pump excitation along a well defined transition dipole induces a loss of isotropy in the molecular target, that breaks the $\varphi$ cylindrical symmetry required to retrieve the $ARPES$ distribution from its VMI 2D-projection $PROJ$. Nevertheless, in fenchone, this cylindrical symmetry is almost restored in the `difference'-image, defined as $PROJ^{(+1,e)}(\theta,E_{kin})-PROJ^{(-1,e)}(\theta,E_{kin})$, obtained by subtracting the images measured with left and right probe polarizations, and called $(LCP-RCP)$ images in Figure 1 (see Supplementary Informations). Fitting this $(LCP-RCP)$ `difference'-image by p-Basex provides the $B_{odd}^{(p,e)}(E_{kin})$ coefficients (see Supplementary Informations).\cite{garcia_two-dimensional_2004} \\
 To extract PECD, which manifests itself as an asymmetry in the photoelectron yield emitted in the forward or backward hemisphere along the optical axis \textbf{\textit{k}} for a given enantiomer and for a given light helicity, normalisation by the PES needs to be done. The normalisation of $B_i$ observables yields the $b_i$ coefficients (Eq. 1). As shown in the SI, the break of cylindrical symmetry prevents access to the `true' $B_0(E_{kin})$ via a p-Basex analysis of the `sum'-image , defined as $PROJ^{(+1,e)}(\theta,E_{kin})+PROJ^{(-1,e)}(\theta,E_{kin})$ and called $(LCP+RCP)$ images in Figure 1. Nevertheless we will still use a p-Basex analysis on this $(LCP+RCP)$ `sum'-image to obtain $\widetilde{B}_0(E_{kin})$, and we use this coefficient to normalise the $B_{odd}^{(p,e)}(E_{kin})$ coefficients from which we calculate a circular dichroism of the 2D photoelectron projection. The quantity we calculate, $\widetilde{\text{PECD}}$, is thus not strictly the same as the PECD usually defined in the litterature and that have been all recorded without a break of cylindrical symmetry. For (1R)-(-)-fenchone ($e=-1$), these definitions are :

\begin{align}
\text{PECD}\left ( E_{kin},e=-1 \right ) &=\frac{\int^{\pi}_0 \left ( ARPES^{(1,-1)} ( \theta,E_{kin} )-ARPES^{(-1,-1)} ( \theta,E_{kin}) \right ) d\theta}{PES\left ( E_{kin}\right )}  \nonumber \\
\text{where} \quad \text{PES}\left(E_{kin}\right) =&\frac{\int^{\pi}_0 \left(ARPES^{(1,-1)}(\theta,E_{kin})+ARPES^{(-1,-1)}(\theta,E_{kin})\right)}{2} d\theta =B_0(E_{kin})\nonumber \\
\text{This yields, in terms of Le} &\text{gendre coefficients :}  \nonumber \\
\text{PECD}\left(E_{kin},e=-1\right)& =-\text{PECD}\left(E_{kin},e=+1\right)=\frac{2B_{1}^{(e=-1)}(E_{kin})-0.5 B_{3}^{(e=-1)}(E_{kin})}{B_0(E_{kin})}  \nonumber \\
\text{with}  \quad B_{odd}^{(e)}&=B_{odd}^{(p=1,e)}=-B_{odd}^{(p=-1,e)}  \nonumber \\
\text{With the cylindrical symme}&\text{try breaking :} & \\
\widetilde{\text{PECD}}\left(E_{kin},-1\right) &=\frac{2B_{1}^{(-1)}(E_{kin})-0.5B_{3}^{(-1)}(E_{kin})}{\tilde{B}_0(E_{kin})}=2\tilde{b}_{1}^{(-1)}(E_{kin})-0.5\tilde{b}_{3}^{(-1)}(E_{kin}) \nonumber
\end{align}

Figure 1 shows the `difference'- and `sum'-images measured at 1, 4 and 8 ps pump-probe delay. The time-and energy-dependences of the $B_{odd}^{(-1)}$ and $\tilde{B}_{even}$ coefficients are extracted from these images. A significant forward PECD, (i.e.) preferential emission of the photoelectron in the \textbf{\textit{k}} direction, is observed on the `difference'-images, demonstrating that loosely bound electronic excited states in chiral molecules can lead to strong asymmetric photoionisation when driven by CPL. We also see that the `sum'-images show a preferential emission along the pump polarization. In order to obtain a time-dependence of this anisotropy, we use also the $\tilde{B}_2 (E_{kin})$ coefficient extracted from the p-Basex analysis of these `sum'-images. Similarly to the $\tilde b_{odd}$ coefficients, normalisation of $\tilde{B}_2$ by $\tilde{B}_0$ is required to obtain $\tilde{b}_2$. Note that $\tilde{b}_2$ coefficient cannot be quantitatively used to evaluate the anisotropy of the $ARPES$, but it will show the same temporal evolution. The $\tilde{B}_4$ coefficient in these `sum'-images is zero, whatever the delay, as expected for a (1+1') resonant photoionisation.\\

\begin{figure}
\includegraphics{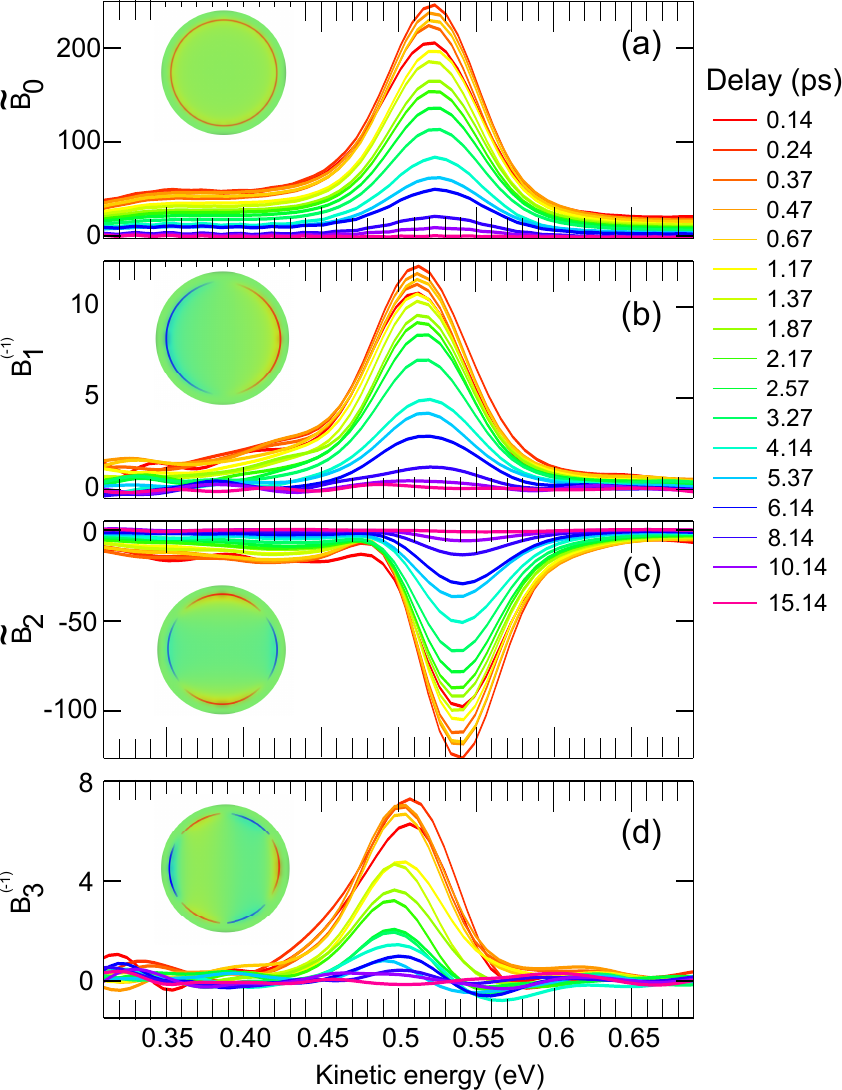}
\caption{The unnormalised Legendre coefficients $B_{odd}^{(-1)}$ and $\tilde{B}_{even}$ as a function of the photoelectron kinetic energy and the pump-probe delay in  (1R)-(-)-fenchone.}
\end{figure}

Both enantiomers have been studied independently and the sign of the $B_{odd}^{(+1)}$ of (1S)-(+) has been reversed to be compared to the (1R)-(-) ones. They are both positive for (1R)-(-)  (negative for (1S)-(+)). Figure 2 shows the energy dependences of the $B_{odd}^{(-1)}$ and $\tilde{B}_{even}$ for different pump-probe delays. The photoelectrons are emitted with a kinetic energy of 0.52 eV irrespective of the pump-probe delay. This 0.52 eV corresponds to the energy expected at the vertical ionisation threshold ($h\nu_{pump}+h\nu_{probe}-IP_{v}=$9.24-8.72 eV -see SI). 

Figure 3 shows the time-dependences of $\tilde{b}_{odd}^{(-1)}$ and  $\tilde{b}_2$ coefficients, extracted from Figure 2 by averaging the $B_i(E_{kin})$ around $E_{kin}$=0.52 eV over 30 meV and by normalising these averages by $\tilde{B}_0(E_{kin})$ treated the same way. The $\tilde{B}_0$ coefficient, proportional to the total number of photoelectron, decays in 3.3 ps while the $\tilde{b}_2$ anisotropy slightly increases in magnitude with a faster rising time, around 2 ps. The asymmetric parameters $\tilde{b}_1^{(-1)}$ and $\tilde{b}_3^{(-1)}$ show opposite trends, with a slow increase of $\tilde{b}_1^{(-1)}$ over 3.2 ps while $\tilde{b}_3^{(-1)}$ decreases faster in only 1.4 ps. Using Eq. 1, the $\widetilde{\text{PECD}}$ has the same sign and a similar magnitude, to that observed at that same total energy in single-photon VUV photoionisation (9.3 eV)\cite{nahon_determination_2016} even though the ionisation process is different. In the experiments of ref. 10, the outgoing electron was scattered from the HOMO orbital in the ground electronic state potential, while here, the outgoing electron is emitted from the 3s Rydberg wavefunction and scatters further from the ionic core. \\
\begin{figure}
\includegraphics{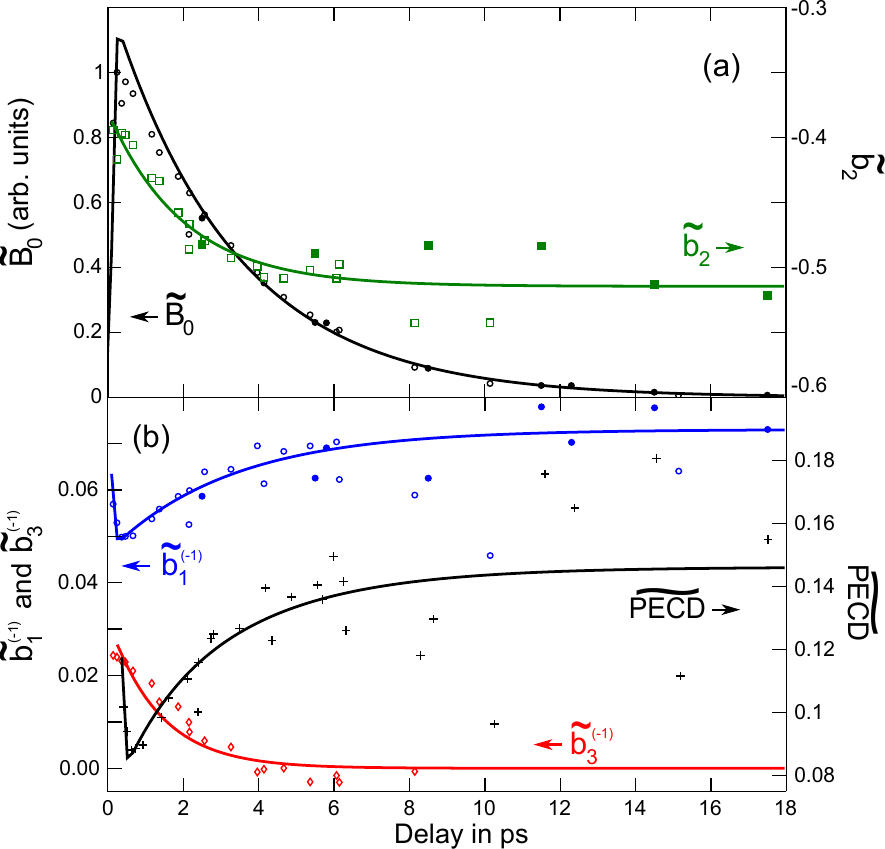}
\caption{Temporal evolution of $\tilde{b}_{odd}^{(-1)}$, $\tilde{b}_{2}$ and $\tilde{b}_{0}$ with empty markers data recorded on  (1R)-(-) enantiomer and filled ones on  (1S)-(+) enantiomer (with a sign inversion on the $\tilde{b}_{odd}^{(+1)}$). (a) $\tilde{b}_{0}$ in black, $\tilde{b}_{2}$ in green, (b) $\tilde{b}_{1}^{(-1)}$ in blue, $\tilde{b}_{3}^{(-1)}$ in red and $\widetilde{\text{PECD}}$ in black. Table 1 summarizes the time constants extracted from the fits (see SI). When the population in 3s state starts to be less than 10\% (delay$\geq$7ps), the normalisation by a small $\tilde{B}_{0}$ increases the dispersion of the normalised coefficients.}
\end{figure}

What are the dynamics revealed by these different timescales ? The absorption of the pump pulse creates a vibrational wavepacket in the 3s Rydberg state. This wavepacket will undergo vibrational relaxation dynamics towards the equilibrium geometry of the 3s state as illustrated in Figure 4(a). In addition, vibronic couplings may occur, leading to internal conversion towards either the lowest valence state (S$_1$) or the ground state (S$_0$) as shown in Figure 4(c). Nothing is known about these two ultrafast processes. In the present excitation scheme, the vibrational energy in both S$_1$ ($\sim$1.6 eV) and S$_0$ ($\sim$6 eV) is too high to get significant Franck-Condon factors at only $\sim$0.5 eV above the IP$_v$. The probe energy is thus not high enough to photoionise molecules vibrationnally excited in S$_0$ or in S$_1$. \\

\begin{table}[h]
\begin{tabular}{ccc}
\toprule
Observables & Time constant in ps & Converging value of the $\tilde{b}_i^{(-1)}$ fit at long delay\\
\midrule
$\tilde{B}_0$ & 3.28$\pm$0.05 & 0 \\
$\tilde{b}_1^{(-1)}$ & 3.25$\pm$0.5 & 0.073$\pm$0.017 \\
$\tilde{b}_2$ & 2.0$\pm$0.3 & -\\
$\tilde{b}_3^{(-1)}$ & 1.4$\pm$0.3 & 0 \\
\bottomrule
\end{tabular}
\caption{Time constants extracted from the fits of the normalised Legendre polynomial coefficients $\tilde{b}_i^{(-1)}$ and  $\tilde{b}_{2}$ dependences shown in Figure 3. The uncertainties correspond to a 50\% confidence bounds.}
\end{table}


Consequently, the 3.28 ps $\tilde{B}_0$ decay mirrors the lifetime of the 3s Rydberg state and the internal conversion timescale from 3s to the lower valence states. The $\tilde{b}_1$ coefficient is mostly determined by the Rydberg-electron continuum transition dipole moment while $\tilde{b}_3$ being a higher order term, will depend on the anisotropy of excitation created by the pump pulse. Similarly a $\beta$ time-dependence within a given electronic state is, in general, also associated to a decay of the anisotropy of excitation created by the pump.\cite{tsubouchi_photoelectron_2001, stolow_time-resolved_2008} The calculations presented in the SI predict that the 3s-S$_0$ transition dipole moment is aligned along the C1-C3 axis, as shown on Figure 4(a). The free rotations of the excited molecules induce a loss of the anisotropy of excitation, over a timescale defined by the moments of inertia of the molecules and the rotational temperature of the gas.\cite{blokhin_dynamics_2003}
The moments of inertia of the asymmetrical top fenchone have been calculated from the S$_0$ geometry to I$_{x,y,z}$=1159, 1555 and 1876 amu.bohr$^2$ with the axis represented on Figure 4(a). The decay time due to the anisotropy loss is typically 1/3 of the dephasing time of the rotational periods $\sqrt{\left(I_y/(kT_{rot})\right)}$, i.e. between 1.4 and 2.4 ps for $T_{rot}$ $\sim$10-30 K as expected for an Even-Lavie pulsed valve source.\cite{blokhin_dynamics_2003,horke_time-resolved_2015} 
 This is consistent with the 1.4 ps decay measured on $\tilde{b}_3^{(-1)}$ and the 2 ps on $\tilde{b}_2$, confirming the sensitivity of these two observables to the average laboratory-frame alignment created by the pump transition. Here, the 3s-S$_0$ dipole moment is almost aligned along the I$_y$ rotational axis. To assign the time-dependence of $\tilde{b}_3^{(-1)}$ and $\tilde{b}_2$ due to I$_x$ and I$_z$ rotations with similar periods than I$_y$, is a challenging task beyond the scope of this letter. This would require a systematic study for different relative polarizations of the pump and probe pulses to extract the spherical tensor components of the photoionisation matrix element. \\
Beyond these global picosecond dynamics, two different mechanisms can be inferred from Figure 3 : $\tilde{b}_1^{(-1)}$ is found to reach a minimum, about 0.05, at 400 fs, and later on attains a constant value of  $\sim$ 0.073. Due to the 3s resonance, this $\tilde{b}_1^{(-1)}$ coefficient, appearing in Eq. 1, is seen experimentally to be the dominant contribution to the $\widetilde{\text{PECD}}$. Consequently the same time-dependence is visible on  $\widetilde{\text{PECD}}$ with a final value converging to +14.6 $\pm$ 3.3\% for (1R)-(-) enantiomer (details in the SI). This asymmetry has the same sign and amplitude as that measured with single VUV photon in this same energy range (+15.4 \%)\cite{nahon_determination_2016}. More relevantly, the $\widetilde{\text{PECD}}$ value at $\sim$0 fs delay in this (1+1') ionisation can be also compared to the (2+1) fs-CPL one recorded on fenchone at 398 nm,\cite{lux_circular_2012,Beaulieu_2016}  while the $\tilde{b}_{odd}$ values can not be since the number and polarizations of photons used to reach the resonance are different. 
In both 0fs-(1+1') and (2+1) fs-CPL ionisation, the PECD observed is in the range of +10\%. More experiments are required before one can conclude whether or not intermediate resonances in REMPI fixes quantitatively the PECD.\\

To come back on the 400 fs time transient of $\tilde{b}_1^{(-1)}$, both TD-DFT and SA6-CASSCF(4,7)/aug-cc-pVDZ calculations predict that the vibrational wavepacket created by the pump pulse will be driven by the steepest descent path from the Franck–Condon region to the minimum-energy region of the 3s potential mostly along the stretching of the C1-C2 bond of the six carbon cycle, as illustrated in Figure 4(a). This complex vibrational dynamics is expected to take place with energy flow between different vibrational modes, leading to a larger diversity of populated vibrational modes. PECD measurements on static molecules have revealed a high sensitivity to the vibrational excitation of the cation, with even the possibility of sign inversion, i.e. a flipping of the recorded asymmetry, upon excitation of certain vibrational modes.\cite{garcia_vibrationally_2013,Beaulieu_2016}
In this context, we suggest that the first 400 fs decay observed on $\tilde{b}_1^{(-1)}$ and $\widetilde{\text{PECD}}$ reflects a part of intramolecular vibrational relaxation (IVR) taking place in the 3s Rydberg state. Indeed the low degree of molecular symmetry does not restrain vibrational interaction such that fast IVR can take place. A lowering of the  $\widetilde{\text{PECD}}$ in this short 400 fs window would result from the increasing number of populated vibrational modes, with a priori different contributions to $\tilde{b}_1^{(-1)}$, both in magnitude and sign, relative to the vibrational wavepacket created at 0 fs and thus leading to a blurring of the overall photoelectron asymmetry. This vibrational relaxation would involve, within the 0.25 eV kinetic energy available in the 3s state, principally the low frequency, (i.e) large amplitude, vibrational modes for example the bending of bicyclic skeleton, twisting modes and methyl torsion, all of which are characterized by large anharmonic couplings. Figure 4(b) shows the vibrational quanta lower than 1000 cm$^{-1}$ (0.12 eV), while Figure 4(d) shows the vibrational mode with the lowest frequency (at 84 cm$^{-1}$). The 3s internal conversion onto lower valence states occurs however in a subset of the available degrees of freedom, meaning that all the vibrational modes populated directly or by IVR in the 3s state during the first 400 fs do not relax at the same time.\cite{solling_non-ergodic_2014,wende_population-controlled_2014} At long delay, we expect that only the less efficiently coupled vibrational modes, the spectator modes of the radiationless transition, will remain populated. It seems that in fenchone, these spectator modes are related to a larger PECD as shown by the rising of the $\tilde{b}_1^{(-1)}$ over $\sim$ 3.25 ps.

\begin{figure}
\includegraphics[width=8 cm]{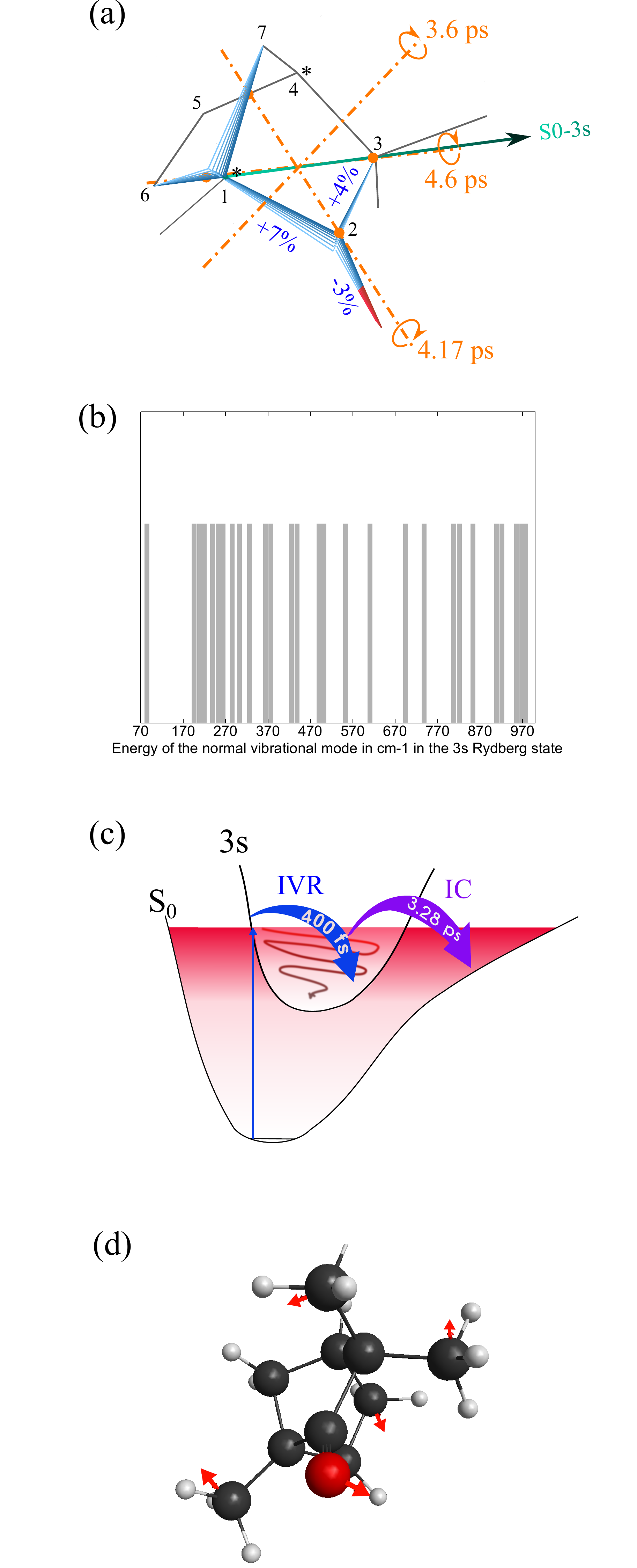}
\caption{(a) Change in the nuclear structure of fenchone from the ground electronic state to the equilibrium geometry of the 3s Rydberg state. The most important relative changes of bond lengths are indicated. The arrow shows the direction of the 3s-S$_0$ transition dipole moment and the orange axis, the rotational axis with their rotational dephasing time for 30 K rotational temperature. (b) The different vibrational quantum of  fenchone in the 3s state below 1000 cm$^{-1}$. (c) The relaxation processes revealed in the TR-PECD. (d) The lowest vibrational mode with a quantum of 84 cm$^{-1}$.}
\end{figure}

This interpretation requires  theoretical support with calculations of TR-PECD that include vibrations, this is far beyond the present state-of-the-art. In particular, our results indicate that PECD would be very sensitive to the initial vibrational states, while previous investigations only considered the effect of vibrational excitation of the final cationic state.\cite{garcia_vibrationally_2013, powis_communication_2014}\\
The $\widetilde{\text{PECD}}$ from the 3s Rydberg state ($\sim$14.6 $\pm$ 3.3\%) is as strong as the PECD recorded from the ground electronic state (15.4\%). The $\widetilde{\text{PECD}}$ magnitude obtained at large delays is rather impressive, considering that the 3s outgoing Rydberg electron scatters less often in the inner part of the molecular potential but more often at distance further from the chiral carbons compared to an electron outgoing from the HOMO ground state orbital. This shows that the phases of partial wavefunctions can be determined by their values far  from the ionic core and that PECD is thus a very long range probe of the whole molecular potential, as already observed in one-photon PECD, for instance in clusters.\cite{nahon_effects_2010,powis_photoionisation_2013} At the same time, by using a pump step, partial alignment is achieved and by exciting a Rydberg state, a smaller number of angular momentum components are required to describe the photoelectron continuum. It would be interesting to investigate these two latter conditions to determine their respective impact on the large asymmetry that is observed here.\\
In conclusion, we have shown that a large $\widetilde{\text{PECD}}$ ($>$10\%) could be observed in photoionisation from excited chiral molecules even if the outgoing electron is emitted from an extended wavefunction such as a Rydberg state. Furthermore, we have demonstrated that Time-Resolved PECD (TR-PECD) could be used as a sensitive probe of ultrafast dynamics in chiral molecules with femtosecond resolution. The different Legendre coefficients reveal dynamics related to alignment ($\tilde{b}_3^{(-1)}$ ), vibrational relaxation ($\tilde{b}_1^{(-1)}$ ) as well as electronic conversion ($\tilde{B}_0$), all behave independently with different time constants. Only the $\tilde{b}_1^{(-1)}$ observable appears here sensitive to sub-ps dynamics such as IVR. \\
In particular, odd coefficients, such as $\tilde{b}_1$ which is fully proportional to the sine of dephasing between adjacent outgoing waves,\cite{powis_photoelectron_2008} are much more sensitive to the whole molecular potential than even parameters such as $\tilde{B}_0$. TR-PECD provides new and richer opportunities to template molecular dynamics in chiral systems in comparison to previous time-resolved Angle-Resolved PES on achiral samples involving only PES and the classical $\beta$ anisotropy parameter. The ability to probe complex ultrafast relaxation in chiral systems opens up concrete possibilities for the investigation of molecular dynamics such as transient chirality or nuclear motions involved in chiral recognition. These results present a stepping stone for time-resolved chirality and more generally to establish a femtochemistry based on circular polarizations.

\section{Experimental and computational methods}

An Even-Lavie valve is used as a pulsed enantiopure fenchone source with helium as carrier gas to avoid cluster formation. The experiment was performed using the Aurore laser system at CELIA, which delivers 20 mJ, 25 fs, 800 nm pulses at 1 kHz. Only 2 mJ is used in the present experiment. The presented results are obtained by scanning the pump-probe delays typically 16 times and at each delay, one helicity image is a record over 45000 laser shots. For each scan, two background images corresponding to the probe pulse alone in each helicity are recorded and subtracted to each image of each delay for a given helicity. This background subtraction is required for the longest pump-probe delay when the signal reaches zero (see Figure S4 of the SI). 
High-level quantum chemistry calculations, using Gaussian 09, Molpro and GAMESS-US quantum chemistry packages, have been performed on the ground and first five electronic excited states of fenchone in order to help in the interpretation of the experimental data.  All the computational details and results can be found in Supporting Information.

\begin{suppinfo}

Details of the calculations of the geometry and energies of the excited states, the experimental set-up, the adiabatic ionisation potential deduced experimentally by the energy balance of the PES, the effect of breaking cylindrical symmetry and the fitting procedures used in Figure 3.

\end{suppinfo}

\section{Author Information}
The authors declare no competing financial interests.

\begin{acknowledgement}

This work was financially supported by Agence Nationale de la Recherche (ANR) (ANR-14-CE32-0014 MISFITS) and Universit\'e of Bordeaux. This project has received funding from the European Research Council (ERC) under the European Union's Horizon 2020 research and innovation programme n$^\circ$682978 - EXCITERS.
 The authors thank R. Bouillaud and L. Merzeau for their technical assistance;  E. M\'evel and E. Constant for providing key apparatus to the experiment, and Gustavo Garcia for stimulating discussions. SB acknowledges the support of a NSERC Vanier Canada Graduate Scholarship.

\end{acknowledgement}

\bibliography{bib_JPCL}

\providecommand{\latin}[1]{#1}
\makeatletter
\providecommand{\doi}
  {\begingroup\let\do\@makeother\dospecials
  \catcode`\{=1 \catcode`\}=2\doi@aux}
\providecommand{\doi@aux}[1]{\endgroup\texttt{#1}}
\makeatother
\providecommand*\mcitethebibliography{\thebibliography}
\csname @ifundefined\endcsname{endmcitethebibliography}
  {\let\endmcitethebibliography\endthebibliography}{}
\begin{mcitethebibliography}{26}
\providecommand*\natexlab[1]{#1}
\providecommand*\mciteSetBstSublistMode[1]{}
\providecommand*\mciteSetBstMaxWidthForm[2]{}
\providecommand*\mciteBstWouldAddEndPuncttrue
  {\def\EndOfBibitem{\unskip.}}
\providecommand*\mciteBstWouldAddEndPunctfalse
  {\let\EndOfBibitem\relax}
\providecommand*\mciteSetBstMidEndSepPunct[3]{}
\providecommand*\mciteSetBstSublistLabelBeginEnd[3]{}
\providecommand*\EndOfBibitem{}
\mciteSetBstSublistMode{f}
\mciteSetBstMaxWidthForm{subitem}{(\alph{mcitesubitemcount})}
\mciteSetBstSublistLabelBeginEnd
  {\mcitemaxwidthsubitemform\space}
  {\relax}
  {\relax}

\bibitem[Zehnacker and Suhm(2008)Zehnacker, and Suhm]{zehnacker_chirality_2008}
Zehnacker,~A.; Suhm,~M. Chirality recognition between neutral molecules in the
  gas phase. \emph{Angewandte Chemie International Edition} \textbf{2008},
  \emph{47}, 6970--6992\relax
\mciteBstWouldAddEndPuncttrue
\mciteSetBstMidEndSepPunct{\mcitedefaultmidpunct}
{\mcitedefaultendpunct}{\mcitedefaultseppunct}\relax
\EndOfBibitem
\bibitem[Berova \latin{et~al.}(2000)Berova, Nakanishi, and
  Woody]{berova_circular_2000}
Berova,~N., Nakanishi,~K., Woody,~R., Eds. \emph{Circular dichroism: principles
  and applications}, 2nd ed.; Wiley-VCH: New York, 2000\relax
\mciteBstWouldAddEndPuncttrue
\mciteSetBstMidEndSepPunct{\mcitedefaultmidpunct}
{\mcitedefaultendpunct}{\mcitedefaultseppunct}\relax
\EndOfBibitem
\bibitem[Ritchie(1975)]{ritchie_theoretical_1975}
Ritchie,~B. Theoretical studies in photoelectron spectroscopy. {Molecular}
  optical activity in the region of continuous absorption and its
  characterization by the angular distribution of photoelectrons.
  \emph{Physical Review A} \textbf{1975}, \emph{12}, 567--574\relax
\mciteBstWouldAddEndPuncttrue
\mciteSetBstMidEndSepPunct{\mcitedefaultmidpunct}
{\mcitedefaultendpunct}{\mcitedefaultseppunct}\relax
\EndOfBibitem
\bibitem[Powis(2000)]{powis_photoelectron_2000}
Powis,~I. Photoelectron circular dichroism of the randomly oriented chiral
  molecules glyceraldehyde and lactic acid. \emph{The Journal of Chemical
  Physics} \textbf{2000}, \emph{112}, 301--310\relax
\mciteBstWouldAddEndPuncttrue
\mciteSetBstMidEndSepPunct{\mcitedefaultmidpunct}
{\mcitedefaultendpunct}{\mcitedefaultseppunct}\relax
\EndOfBibitem
\bibitem[B{\"o}wering \latin{et~al.}(2001)B{\"o}wering, Lischke, Schmidtke,
  M{\"u}ller, Khalil, and Heinzmann]{bowering_asymmetry_2001}
B{\"o}wering,~N.; Lischke,~T.; Schmidtke,~B.; M{\"u}ller,~N.; Khalil,~T.;
  Heinzmann,~U. Asymmetry in photoelectron emission from chiral molecules
  induced by circularly polarized light. \emph{Physical Review Letters}
  \textbf{2001}, \emph{86}, 1187--1190\relax
\mciteBstWouldAddEndPuncttrue
\mciteSetBstMidEndSepPunct{\mcitedefaultmidpunct}
{\mcitedefaultendpunct}{\mcitedefaultseppunct}\relax
\EndOfBibitem
\bibitem[Nahon \latin{et~al.}(2015)Nahon, Garcia, and
  Powis]{nahon_valence_2015}
Nahon,~L.; Garcia,~G.~A.; Powis,~I. Valence shell one-photon photoelectron
  circular dichroism in chiral systems. \emph{Journal of Electron Spectroscopy
  and Related Phenomena} \textbf{2015}, \emph{204}, 322--334\relax
\mciteBstWouldAddEndPuncttrue
\mciteSetBstMidEndSepPunct{\mcitedefaultmidpunct}
{\mcitedefaultendpunct}{\mcitedefaultseppunct}\relax
\EndOfBibitem
\bibitem[Lux \latin{et~al.}(2012)Lux, Wollenhaupt, Bolze, Liang, K{\"o}hler,
  Sarpe, and Baumert]{lux_circular_2012}
Lux,~C.; Wollenhaupt,~M.; Bolze,~T.; Liang,~Q.; K{\"o}hler,~J.; Sarpe,~C.;
  Baumert,~T. Circular dichroism in the photoelectron angular distributions of
  camphor and fenchone from multiphoton ionization with femtosecond laser
  pulses. \emph{Angewandte Chemie International Edition} \textbf{2012},
  \emph{51}, 5001--5005\relax
\mciteBstWouldAddEndPuncttrue
\mciteSetBstMidEndSepPunct{\mcitedefaultmidpunct}
{\mcitedefaultendpunct}{\mcitedefaultseppunct}\relax
\EndOfBibitem
\bibitem[Lehmann \latin{et~al.}(2013)Lehmann, Ram, Powis, and
  Janssen]{lehmann_imaging_2013}
Lehmann,~C.~S.; Ram,~N.~B.; Powis,~I.; Janssen,~M. H.~M. Imaging photoelectron
  circular dichroism of chiral molecules by femtosecond multiphoton coincidence
  detection. \emph{The Journal of Chemical Physics} \textbf{2013}, \emph{139},
  234307\relax
\mciteBstWouldAddEndPuncttrue
\mciteSetBstMidEndSepPunct{\mcitedefaultmidpunct}
{\mcitedefaultendpunct}{\mcitedefaultseppunct}\relax
\EndOfBibitem
\bibitem[Beaulieu \latin{et~al.}(2016)Beaulieu, Ferr\'e, G\'eneaux, Canonge,
  Descamps, Fabre, Fedorov, L\'egar\'e, Petit, Ruchon, Blanchet, Mairesse, and
  Pons]{Beaulieu_2016}
Beaulieu,~S.; Ferr\'e,~A.; G\'eneaux,~R.; Canonge,~R.; Descamps,~D.; Fabre,~B.;
  Fedorov,~N.; L\'egar\'e,~F.; Petit,~S.; Ruchon,~T. \latin{et~al.}
  Universality of photoelectron circular dichroism in the photoionization of
  chiral molecules. \emph{New Journal of Physics} \textbf{2016}, \emph{18},
  102002\relax
\mciteBstWouldAddEndPuncttrue
\mciteSetBstMidEndSepPunct{\mcitedefaultmidpunct}
{\mcitedefaultendpunct}{\mcitedefaultseppunct}\relax
\EndOfBibitem
\bibitem[Nahon \latin{et~al.}(2016)Nahon, Nag, Garcia, Myrgorodska,
  Meierhenrich, Beaulieu, Wanie, Blanchet, G\'eneaux, and
  Powis]{nahon_determination_2016}
Nahon,~L.; Nag,~L.; Garcia,~G.; Myrgorodska,~I.; Meierhenrich,~U.~J.;
  Beaulieu,~S.; Wanie,~V.; Blanchet,~V.; G\'eneaux,~R.; Powis,~I. Determination
  of accurate electron chiral asymmetries in fenchone and camphor in the {VUV}
  range: sensitivity to isomerism and enantiomeric purity. \emph{Phys. Chem.
  Chem. Phys.} \textbf{2016}, \emph{18}, 12696--12706\relax
\mciteBstWouldAddEndPuncttrue
\mciteSetBstMidEndSepPunct{\mcitedefaultmidpunct}
{\mcitedefaultendpunct}{\mcitedefaultseppunct}\relax
\EndOfBibitem
\bibitem[Ferré \latin{et~al.}(2015)Ferré, Handschin, Dumergue, Burgy, Comby,
  Descamps, Fabre, Garcia, Géneaux, Merceron, Mével, Nahon, Petit, Pons,
  Staedter, Weber, Ruchon, Blanchet, and Mairesse]{ferre_table-top_2015}
Ferré,~A.; Handschin,~C.; Dumergue,~M.; Burgy,~F.; Comby,~A.; Descamps,~D.;
  Fabre,~B.; Garcia,~G.~A.; Géneaux,~R.; Merceron,~L. \latin{et~al.}  A
  table-top ultrashort light source in the extreme ultraviolet for circular
  dichroism experiments. \emph{Nature Photonics} \textbf{2015}, \emph{9},
  93--98\relax
\mciteBstWouldAddEndPuncttrue
\mciteSetBstMidEndSepPunct{\mcitedefaultmidpunct}
{\mcitedefaultendpunct}{\mcitedefaultseppunct}\relax
\EndOfBibitem
\bibitem[Lux \latin{et~al.}(2015)Lux, Wollenhaupt, Sarpe, and
  Baumert]{lux_photoelectron_2015}
Lux,~C.; Wollenhaupt,~M.; Sarpe,~C.; Baumert,~T. Photoelectron circular
  dichroism of bicyclic ketones from multiphoton ionization with femtosecond
  lLaser pulses. \emph{ChemPhysChem} \textbf{2015}, \emph{16}, 115--137\relax
\mciteBstWouldAddEndPuncttrue
\mciteSetBstMidEndSepPunct{\mcitedefaultmidpunct}
{\mcitedefaultendpunct}{\mcitedefaultseppunct}\relax
\EndOfBibitem
\bibitem[Pulm \latin{et~al.}(1997)Pulm, Schramm, Hormes, Grimme, and
  Peyerimhoff]{pulm_theoretical_1997}
Pulm,~F.; Schramm,~J.; Hormes,~J.; Grimme,~S.; Peyerimhoff,~S. Theoretical and
  experimental investigations of the electronic circular dichroism and
  absorption spectra of bicyclic ketones. \emph{Chemical Physics}
  \textbf{1997}, \emph{224}, 143--155\relax
\mciteBstWouldAddEndPuncttrue
\mciteSetBstMidEndSepPunct{\mcitedefaultmidpunct}
{\mcitedefaultendpunct}{\mcitedefaultseppunct}\relax
\EndOfBibitem
\bibitem[Garcia \latin{et~al.}(2004)Garcia, Nahon, and
  Powis]{garcia_two-dimensional_2004}
Garcia,~G.~A.; Nahon,~L.; Powis,~I. Two-dimensional charged particle image
  inversion using a polar basis function expansion. \emph{Review of Scientific
  Instruments} \textbf{2004}, \emph{75}, 4989--4996\relax
\mciteBstWouldAddEndPuncttrue
\mciteSetBstMidEndSepPunct{\mcitedefaultmidpunct}
{\mcitedefaultendpunct}{\mcitedefaultseppunct}\relax
\EndOfBibitem
\bibitem[Tsubouchi \latin{et~al.}(2001)Tsubouchi, Whitaker, Wang, Kohguchi, and
  Suzuki]{tsubouchi_photoelectron_2001}
Tsubouchi,~M.; Whitaker,~B.; Wang,~L.; Kohguchi,~H.; Suzuki,~T. Photoelectron
  imaging on time-dependent molecular alignment created by a femtosecond laser
  pulse. \emph{Physical Review Letters} \textbf{2001}, \emph{86},
  4500--4503\relax
\mciteBstWouldAddEndPuncttrue
\mciteSetBstMidEndSepPunct{\mcitedefaultmidpunct}
{\mcitedefaultendpunct}{\mcitedefaultseppunct}\relax
\EndOfBibitem
\bibitem[Stolow and Underwood(2008)Stolow, and
  Underwood]{stolow_time-resolved_2008}
Stolow,~A.; Underwood,~J. Time-resolved photoelectron spectroscopy of
  nonadiabatic dynamics in polyatomic molecules. \emph{Advances in Chemical
  Physics, Vol 139} \textbf{2008}, \emph{139}, 497--583\relax
\mciteBstWouldAddEndPuncttrue
\mciteSetBstMidEndSepPunct{\mcitedefaultmidpunct}
{\mcitedefaultendpunct}{\mcitedefaultseppunct}\relax
\EndOfBibitem
\bibitem[Blokhin \latin{et~al.}(2003)Blokhin, Gelin, Khoroshilov, Kryukov, and
  Sharkov]{blokhin_dynamics_2003}
Blokhin,~A.; Gelin,~M.; Khoroshilov,~E.; Kryukov,~I.; Sharkov,~A. Dynamics of
  optically induced anisotropy in an ensemble of asymmetric top molecules in
  the gas phase. \emph{Optics and Spectroscopy} \textbf{2003}, \emph{95},
  346--352\relax
\mciteBstWouldAddEndPuncttrue
\mciteSetBstMidEndSepPunct{\mcitedefaultmidpunct}
{\mcitedefaultendpunct}{\mcitedefaultseppunct}\relax
\EndOfBibitem
\bibitem[Horke \latin{et~al.}(2015)Horke, Chatterley, Bull, and
  Verlet]{horke_time-resolved_2015}
Horke,~D.~A.; Chatterley,~A.~S.; Bull,~J.~N.; Verlet,~J. R.~R. Time-resolved
  photodetachment anisotropy: gas-phase rotational and vibrational dynamics of
  the fluorescein anion. \emph{The Journal of Physical Chemistry Letters}
  \textbf{2015}, \emph{6}, 189--194\relax
\mciteBstWouldAddEndPuncttrue
\mciteSetBstMidEndSepPunct{\mcitedefaultmidpunct}
{\mcitedefaultendpunct}{\mcitedefaultseppunct}\relax
\EndOfBibitem
\bibitem[Garcia \latin{et~al.}(2013)Garcia, Nahon, Daly, and
  Powis]{garcia_vibrationally_2013}
Garcia,~G.~A.; Nahon,~L.; Daly,~S.; Powis,~I. Vibrationally induced inversion
  of photoelectron forward-backward asymmetry in chiral molecule
  photoionization by circularly polarized light. \emph{Nature Communications}
  \textbf{2013}, \emph{4}\relax
\mciteBstWouldAddEndPuncttrue
\mciteSetBstMidEndSepPunct{\mcitedefaultmidpunct}
{\mcitedefaultendpunct}{\mcitedefaultseppunct}\relax
\EndOfBibitem
\bibitem[S{\v{o}}lling \latin{et~al.}(2014)S{\v{o}}lling, Kuhlman, Stephansen,
  Klein, and M{\v{o}}ller]{solling_non-ergodic_2014}
S{\v{o}}lling,~T.~I.; Kuhlman,~T.~S.; Stephansen,~A.~B.; Klein,~L.~B.;
  M{\v{o}}ller,~K.~B. The non-ergodic nature of internal conversion.
  \emph{ChemPhysChem} \textbf{2014}, \emph{15}, 249--259\relax
\mciteBstWouldAddEndPuncttrue
\mciteSetBstMidEndSepPunct{\mcitedefaultmidpunct}
{\mcitedefaultendpunct}{\mcitedefaultseppunct}\relax
\EndOfBibitem
\bibitem[Wende \latin{et~al.}(2014)Wende, Liebel, Schnedermann, Pethick, and
  Kukura]{wende_population-controlled_2014}
Wende,~T.; Liebel,~M.; Schnedermann,~C.; Pethick,~R.~J.; Kukura,~P.
  Population-controlled impulsive vibrational spectroscopy: background- and
  baseline-free Raman spectroscopy of excited electronic states. \emph{The
  Journal of Physical Chemistry A} \textbf{2014}, \emph{118}, 9976--9984\relax
\mciteBstWouldAddEndPuncttrue
\mciteSetBstMidEndSepPunct{\mcitedefaultmidpunct}
{\mcitedefaultendpunct}{\mcitedefaultseppunct}\relax
\EndOfBibitem
\bibitem[Powis(2014)]{powis_communication_2014}
Powis,~I. Communication: {The} influence of vibrational parity in chiral
  photoionization dynamics. \emph{The Journal of Chemical Physics}
  \textbf{2014}, \emph{140}, 111103\relax
\mciteBstWouldAddEndPuncttrue
\mciteSetBstMidEndSepPunct{\mcitedefaultmidpunct}
{\mcitedefaultendpunct}{\mcitedefaultseppunct}\relax
\EndOfBibitem
\bibitem[Nahon \latin{et~al.}(2010)Nahon, Garcia, Soldi-Lose, Daly, and
  Powis]{nahon_effects_2010}
Nahon,~L.; Garcia,~G.~A.; Soldi-Lose,~H.; Daly,~S.; Powis,~I. Effects of
  dimerization on the photoelectron angular distribution parameters from chiral
  camphor enantiomers obtained with circularly polarized vacuum-ultraviolet
  radiation. \emph{Physical Review A} \textbf{2010}, \emph{82},
  032514--10\relax
\mciteBstWouldAddEndPuncttrue
\mciteSetBstMidEndSepPunct{\mcitedefaultmidpunct}
{\mcitedefaultendpunct}{\mcitedefaultseppunct}\relax
\EndOfBibitem
\bibitem[Powis \latin{et~al.}(2013)Powis, Daly, Tia, Miranda, Garcia, and
  Nahon]{powis_photoionisation_2013}
Powis,~I.; Daly,~S.; Tia,~M.; Miranda,~B. C.~d.; Garcia,~G.~A.; Nahon,~L. A
  photoionization investigation of small, homochiral clusters of glycidol using
  circularly polarized radiation and velocity map electron–ion coincidence
  imaging. \emph{Physical Chemistry Chemical Physics} \textbf{2013}, \emph{16},
  467--476\relax
\mciteBstWouldAddEndPuncttrue
\mciteSetBstMidEndSepPunct{\mcitedefaultmidpunct}
{\mcitedefaultendpunct}{\mcitedefaultseppunct}\relax
\EndOfBibitem
\bibitem[Powis(2008)]{powis_photoelectron_2008}
Powis,~I. Photoelectron circular dichroism in chiral molecules. \emph{Advances
  in Chemical Physics} \textbf{2008}, 267--329\relax
\mciteBstWouldAddEndPuncttrue
\mciteSetBstMidEndSepPunct{\mcitedefaultmidpunct}
{\mcitedefaultendpunct}{\mcitedefaultseppunct}\relax
\EndOfBibitem
\end{mcitethebibliography}

\end{document}